\documentclass[12pt,a4paper]{article}
\usepackage[utf8]{inputenc}
\usepackage[T1]{fontenc}

\usepackage{amsmath}
\usepackage{amssymb}
\usepackage{tikz}
\usepackage{tabularx}
\usepackage{authblk}
\usepackage[title]{appendix}
\usepackage{graphicx}
\usepackage{color} 
\usepackage[linktoc=page]{hyperref}
\hypersetup{
	colorlinks=true,
	linkcolor=blue,
	urlcolor=blue,
	bookmarks=true,
	citecolor=[rgb]{0.54,0,0}
}
\author{Yoav Zigdon}
\affil{{\normalsize \textit{School of Physics and Astronomy, Tel Aviv University, Ramat Aviv, 69978, Israel}} \\
	{\normalsize yoavzi(at)tauex.tau.ac.il}}
\date{}
\title{Bridging Worldsheet CFTs and Wormholes}
\begin{document}
	\maketitle
	\begin{abstract}		
		 I provide multiple examples of conformal field theories (CFTs) on the worldsheet that describe string propagation in target space wormholes connecting two disjoint asymptotic manifolds. The worldsheet approach goes beyond the framework of supergravity by incorporating wormholes for which the size of the throat is comparable to the string scale. Typically, strongly coupled CFTs describe these stringy wormholes, which include Euclidean wormholes, double cones, and Einstein-Rosen bridges. Finally, I interpret a certain conformal manifold that contains an $\text{SU}(2)_k \times U(1)$ Wess-Zumino-Witten model as mediating a transition between a closed Universe and a wormhole.  
	\end{abstract}
	\newpage
	\tableofcontents
	\section{Introduction}
	\label{sec:Intro}
	Incorporating wormholes in the gravitational path integral has revealed relations between gravitating systems and quantum systems.  An example is the double-cone wormhole, which was used to account for a ``ramp'' of the spectral form factor \cite{Saad:2018bqo}. Furthermore, reference \cite{Saad:2019lba} showed that Jackiw–Teitelboim (JT) gravity with a negative cosmological constant admits a non-gravitational description when including Euclidean wormholes in the gravitational path integral, valid to all orders in the genus expansion, in terms of a matrix integral of random Hamiltonians in a double-scaled limit. An improved matrix integral was suggested to define JT gravity non-perturbatively in~\cite{Johnson:2019eik}. These results expanded the framework of the gauge/string correspondence from single, non-gravitational quantum systems to ensembles of quantum systems. 
	
	Another example is replica wormholes connecting a disjoint union of copies of boundaries, and arise by modifying the conventional replica trick in Euclidean, non-gravitational quantum field theories. Two of the quantum aspects they explain in black hole physics are (a) curves of entanglement entropy of Hawking radiation against time that are consistent with information preservation,  utilizing the island formula \cite{Penington:2019a}-\cite{Almheiri:2019c}. (b) The dimension of the Hilbert space of black hole microstates, which is approximately the exponential of the Bekenstein-Hawking, was argued to be reproduced by considering shells that collapse inside the black hole, and calculating the rank of their Gram matrix in the presence of replica wormholes~\cite{Balasubramanian:2022gmo}. 
	
	On the other hand, the inclusion of Euclidean wormhole contributions in the context of the conventional gauge/string correspondence is incompatible with the factorization of the partition function of a gauge theory defined on the disconnected boundary manifold of the wormholes  \cite{WittenYau},\cite{MaldacenaMaoz}. Possible solutions to this problem are that either the wormhole contributions cancel against other contributions, or one should filter the partition function of the gauge theory in the large-$N$ limit to subtract off its erratic parts~\cite{Liu2025}. Additionally, wormholes connecting a parent Universe to a baby Universe lead to loss of quantum coherence in the parent Universe~\cite{Hawking:1987mz}, unless the dimension of the Hilbert space of baby Universes is unity~\cite{McNamara:2020uza}.  Also, a discussion emerged about the relation between the presence of wormholes and bi-local interactions in the effective action \cite{Arkani-Hamed:2007cpn},\cite{Guo:2021blh}; non-local interactions constitute a qualitative solution to the information paradox by allowing a transmission of quantum information from the black hole interior to Hawking radiation, while preserving the smoothness of the event horizon~\cite{Giddings:2012gc},\cite{Almheiri:2012rt}.    
	 
	Another class of wormholes is Einstein-Rosen bridges, which appear in classical, two-sided, Lorentzian black hole geometries, where the geometry's signature changes depending on whether the region of the Einstein-Rosen bridge is inside or outside the black hole. 
	Such spacetime bridges were conjectured to correspond to quantum entanglement \cite{VanRaamsdonk:2010pw},\cite{Maldacena:2013xja}, although a clear relation between generic spacetime connections and quantum entanglement has not been established.  For example, references \cite{Marolf:2013dba},\cite{Berkooz} showed that, in the context of gauge/gravity duality, two-point functions on the gauge theory side of simple operators in generic entangled states are too small to admit a semiclassical wormhole interpretation. Another counterexample to the claim that a high degree of entanglement always gives rise to a semiclassical geometric connection appeared in the context of string theory on three-dimensional Anti-de Sitter (AdS) times a compact manifold with an AdS length scale smaller than the string scale, where black holes are absent~\cite{Balthazar:2021xeh},\cite{Martinec:2021vpk}. Thus, a true test of the conjecture in question requires a definition of wormholes in the regime where quantum and/or stringy effects are important. 
	
	The bulk of the discussion of wormholes in the literature is decoupled from a worldsheet approach to describe string propagation in target spaces.  Giddings and Strominger wrote a basic motivation to explore this approach in 1989 \cite{Giddings:1989bq}: ``\textit{Ideally one would like to know whether wormholes exist as exact solutions of the full string equations of motion, i.e. as appropriate $c=26$ or 10 conformal field theories. In practice it may be difficult to find such exact solutions}.''  In response, this paper writes several exact worldsheet conformal field theories (CFTs) for  wormholes.  These constructions are useful for defining the concept of wormholes in the stringy regime where the supergravity approximation fails, though a suppression of string loop-corrections to observables requires a string coupling that is everywhere parametrically small. 
	
	  Using a worldsheet formalism for wormhole targets, one can investigate multiple problems, with two among them 
	\begin{itemize}
	 \item \underline{Stability:} Whether stringy wormholes are stable under small perturbations to the vacuum state of the CFT, and in particular to the phenomenon of brane nucleation. The spectrum of a reflection-positive worldsheet CFT$_2$ admits an interpretation in terms of stable perturbations on top of the wormhole, in that the topology of the target space is unchanged\footnote{The total energy of excitations should not surpass the inverse of the string coupling squared to neglect the backreaction effect.}. Other perturbations could decay by leaking into the ends of the wormhole, or modify its topology by splitting it into disconnected target spaces or connecting its ends.  To examine the effects of perturbations on the CFT quantitatively, one should search for stable fixed points of the deformed CFT and extract the resulting target spaces from them. 
	 \item  \underline{Traversability:} Whether a probe string or brane sent from one disconnected part of the boundary of the stringy wormhole can traverse to the other disconnected parts of the boundary. 
	 The transmission coefficient of this scattering gedanken experiment determines if and to what extent the wormhole is traversable.    
		\end{itemize}
		More problems are written at the end of the paper. 
	The structure of the paper is as follows. In section~\ref{sec:ex}, presented are examples of worldsheet CFTs with target spaces (a) $\mathbb{R}\times$ compact manifold in subsection~\ref{sec:sim}, (b) $\mathbb{R}_t\times $ Euclidean AdS$_2$ in subsection~\ref{sec:AdS2}, (c) AdS$_2$ double-cone in a three-dimensional orbifold of AdS$_3$ in subsection~\ref{sec:double}. Further, a description of Einstein-Rosen bridges in a two-dimensional black hole with asymptotically linear dilaton appears in subsection~\ref{sec:ER}, and a conformal manifold that furnishes closed Universe targets and wormhole targets is explained in subsection~\ref{sec:transition}. In section~\ref{sec:final}, a summary and future directions are provided. 
	\section{Examples}
	\label{sec:ex}
	\subsection{CFTs for Euclidean Wormholes}
	\label{sec:sim}
	This subsection points out that string propagation in trivial Euclidean wormhole target spaces of the direct product form $\mathbb{R}\times$ a compact manifold is described by worldsheet CFTs. The string scale $\sqrt{\alpha'}$ is set to one. \\
	The simplest example of a worldsheet theory for string propagation in a wormhole is the cylinder target space, described by a pair of free bosonic fields $\tau$ and $\rho$, with $\tau$ being compact and $\rho$ non-compact. The worldsheet theory can be defined via a path integral formulation with the action 
	\begin{equation}
		S = \frac{1}{2\pi}\int_{\Sigma} d^2 \sigma \sqrt{\gamma} \gamma^{ab}  \Big( \partial_a \rho \partial_b \rho + \partial_a \tau \partial_b \tau +\Phi_0 R\Big)~~,
	\end{equation} 
	where $\sigma$ parametrizes the two-dimensional worldsheet $\Sigma$, $\gamma_{ab}$ are the components of the intrinsic metric on the worldsheet, the indices $a,b$ run over $1,2$, $\Phi_0$ is a constant value of the dilaton, and $R$ is the Ricci scalar of $\Sigma$.
	The boundary of the wormhole is a disconnected pair of copies of a circle. The spectrum of the CFT corresponds to an infinite set of vertex operators creating string excitations on top of the cylinder target space, possibly with winding about the $\tau$-cycle. These can be viewed as perturbations on top of the cylinder, which are stable, in that the underlying target space is unchanged due to their presence. The spectrum is continuous because it is labeled in part by the center-of-mass momentum conjugate to the center-of-mass position of $\rho$. 
	 Observables are suitably integrated correlation functions of the vertex operators. When the periodicity of $\tau$ surpasses the string length scale (otherwise apply a T-duality), string excitations can traverse from one asymptotic boundary of the wormhole to the other asymptotic boundary because their evolution is free. 
	 
	 A slight variation of this theory is given by adding two fermionic fields to supersymmetrize it. 
	Another variant of this theory is obtained by identifying the field $\rho$ with $-\rho$ and compactifying this field, such that a finite interval $I$ appears as a factor of the target space (in other words, an orbifold $\mathbb{S}^1/\mathbb{Z}_2$), supplanting the non-compact line parametrized by the original variable $\rho$. 
	
	One generalization of this theory is achieved by adding more non-interacting, compact bosons to form a torus of $d$-dimensions, such that the target space is $\mathbb{R} \times \mathbb{T}^d$. Now the two  components of the boundary are $\mathbb{T}^d$ at the two ends of $\rho$. Again, the dilaton is constant and can be tuned to make the string coupling weak. Obtaining a supersymmetric version of this system is straightforward, and for the choice $d=9$, one has an exact solution to the classical superstring equations. Such a wormhole solution is stable under all vertex operator perturbations, because of the tachyon-free nature of the spectrum, when the radius $R$ of each cycle satisfies $R>\sqrt{2\alpha'}$. The on-shell, target-space action of the solution vanishes at tree-level in the string coupling  because all derivatives of the NS-NS, NS-R, R-NS, and RR fields are zero:
	\begin{equation}
		I_{\text{classical}}=0~~.
	\end{equation}  
	Replacing $\mathbb{R}$ by its orbifold and/or $\mathbb{T}^d$ by its orbifold still constitutes wormhole target spaces. 
	 
	A slight generalization to an interacting CFT is the \text{SU}(2)$_k$ Wess-Zumino-Witten (WZW) model decoupled from the non-compact boson $\rho$. Throughout this paper, $k$ denotes the bosonic level of WZW models. A functional integral formalism of the theory on the sphere is based on the action
	\begin{equation}
		\label{rhoSU2}
		S = \frac{1}{2\pi}\int_{\Sigma} d^2\sigma \sqrt{\gamma}\gamma^{ab}  \partial_a \rho \partial_b \rho +S_{\text{WZW},\text{SU}(2)}[g]~~,
	\end{equation} 
	where $S_{\text{WZW,}\text{SU}(2)}[g]$ is the action of the $\text{SU}(2)_k$ WZW model
	\begin{align}
		\label{WZW} 
		S_{\text{WZW}}[g] = \frac{k}{4\pi} \int_{\Sigma} d^2 \sigma \sqrt{|\gamma|} \gamma^{ab} ~\text{Tr}\Big(g^{-1} \partial_a g g^{-1} \partial_b g\Big)+\frac{ik}{12\pi} \int_B \text{Tr} \Big(g^{-1} dg \wedge g^{-1} dg \wedge g^{-1} dg\Big)~~.
	\end{align}
	Here, $g$ is a group element of the $\text{SU}(2)$ group manifold, which can be parametrized in terms of Euler angles as 
	\begin{equation}
		\label{para}
	g=e^{i\frac{\gamma}{2}\sigma_z}e^{i\frac{\beta}{2}\sigma_x}e^{i\frac{\alpha}{2}\sigma_z}~~,	
	\end{equation}
	where $\sigma_x,\sigma_z$ are two of the Pauli matrices, and $B$ is a three-dimensional manifold whose boundary is the worldsheet: $\partial B = \Sigma$. 
	The central charge of the CFT, including the boson $\rho$, is
	\begin{equation}
		\label{c}
		c = \frac{3k}{k+2}+1~~.
	\end{equation}
	 When $k$ is large, the resulting target-space is a four-dimensional wormhole with three-dimensional spheres on its disconnected boundary: $\mathbb{R}\times \mathbb{S\textit{}}^3$. For small $k$, it is proposed that this strongly coupled CFT defines a ``stringy wormhole'', in the same formal way that the strongly coupled CFT of the gauged WZW model based on the $\text{SL}(2,\mathbb{R})_k/\text{U}(1)$ quotient defines a ``stringy Euclidean black hole'' for $k=O(1)$  \cite{Witten:1991yr}. In the example $k=4$, the central charge in Eq.~(\ref{c}) is $3$, and taking a direct product with a free CFT of $23$ compact bosons and the $\{b,c\}$ ghost CFT generates a theory free from the Weyl anomaly. One can further project out ghost states to embed the construction into the bosonic string theory. Note that dilaton remains constant in this construction as well.~\footnote{Incidentally, one of the first examples of a ``semi-wormhole'' target-space with an exact CFT description is \cite{Callan:1991dj}, where NS5-branes induce a transverse geometry $\mathbb{R}_{\rho>0}\times \mathbb{S}^3$ with a linear dilaton direction $\rho$ which grows towards the fivebranes. The word ``semi'' means that only one side of the wormhole is described in a string theory context; close to the coincident fivebranes, one loses the description based on the worldsheet on the sphere. A similar CFT describing string propagation on a cosmological target spacetime $\mathbb{R}_t \times \mathbb{S}^3$ has appeared recently in \cite{Chu:2026rle}.}~The presence of a tachyon implies an instability of this wormhole.

	 One can obtain worldsheet descriptions of products of targets consisting of an $\mathbb{R}$ multiplied by any subset of $\text{SU}(2),\mathbb{T}^d$ (and quotients thereof by $\text{U}(1)$ or $\mathbb{Z}_2$). In particular, one can take an order-one level of the $\text{SU}(2)$ model, and/or a $d$-dimensional torus target-space whose volume is comparable to the string length scale to the power $d$, and the worldsheet formalism works in such regimes, with an interpretation of ``stringy wormholes.'' 
	
	Despite the benign nature of these examples, they are trivial in that the wormholes are not warped; they are simple products of $\mathbb{R}$ (or $\mathbb{R}/\mathbb{Z}_2$ or $\mathbb{S}^1/\mathbb{Z}_2$) with compact manifolds. Below, I present less-trivial examples of interacting CFTs whose coupling is of order one in the regime where the throat size of the wormhole target space approximates the string scale. 
	\subsection{CFT for a EAdS$_2$ Wormhole}
	\label{sec:AdS2}
	This subsection reviews a part of reference \cite{Israel:2004vv} by Isra\"el, Kounnas, Orlando and Petropoulos (IKOP), who deformed the SL$(2,\mathbb{R})_k\times U(1)_{k_G}$ WZW model by an exactly marginal deformation constructed from a bi-linear in currents, and showed that for a limiting value of the deformation parameter, a worldsheet CFT for string propagation in a target spacetime that contains $\mathbb{R}_t \times \mathbb{H}_2$ emerges, where $\mathbb{H}_2$ is the Euclidean continuation of AdS in two dimensions and $\mathbb{R}_t$ represents Lorentzian time.  It is noted that the target-space metric of  $\mathbb{H}_2$ in global coordinates admits an interpretation as a ``double-trumpet'' Euclidean wormhole connecting a pair of disjoint boundaries, with a worldsheet CFT description due to IKOP. The construction can be organized in three layers.\\ 
	In the first layer, one starts with the $(1,0)$ supersymmetric $\text{SL}(2,\mathbb{R})_k\times U(1)$ CFT with central charges 
	\begin{equation}
		\label{c2}
		c_R = 6+\frac{6}{k}~~,~~ c_L =  4+\frac{6}{k-2}~.
	\end{equation}
	 and the action that depends on $t,\phi,\rho,\psi^1,\psi^2,\psi^3$ parametrizing $\text{SL}(2,\mathbb{R})$ and $\varphi,\psi_{\varphi}$ parametrizing $U(1)$:
	\begin{align}
		\label{SL2R}
		S_{\text{SL}(2,\mathbb{R})_k\times U(1)_{k_G}}[t,\phi,\rho,\psi^1,\psi^2,\psi^3,\varphi,\psi_{\varphi}] &= \frac{k}{8\pi} \int d^2 z \Big(\partial \rho \bar{\partial} \rho -\partial t \bar{\partial} t + \partial \phi \bar{\partial}\phi -2\sinh(\rho) \partial \phi \bar{\partial} t\Big)\nonumber\\
		&~~+\frac{1}{2\pi} \int d^2 z \Big(\psi^1 \bar{\partial} \psi ^1 + \psi^2 \bar{\partial} \psi ^2 - \psi^3 \bar{\partial} \psi ^3 \Big)\nonumber\\
		& ~~+ \frac{k_G}{4\pi} \int d^2 z~\partial \varphi \bar{\partial} \varphi+\frac{1}{2\pi}\int d^2 z ~ \psi_{\varphi} \bar{\partial} \psi_{\varphi}~.
	\end{align}
	A holomorphic current in the $\text{SL}(2,\mathbb{R})$ WZW model, $J^3$, will play a role below; in terms of the parametrization $(t,\rho,\phi)$, it is given by
	\begin{equation}
		J^3= k\big(\partial t + \sinh(\rho)\partial \phi\big)~~.
	\end{equation}
	In the second layer, one deforms the above WZW model by an asymmetric, elliptic, ``magnetic'' deformation. It is defined through a quadratic form in conserved currents: 
	\begin{equation}
		\delta S_{\text{magnetic}} =\frac{H}{2\pi}\sqrt{\frac{k_G}{k}}\int d^2 z  \big(J^3 + i\psi^1 \psi^2  \big)\bar{\partial} \varphi~~.
	\end{equation}
	Here, $H$ will be referred to as a ``deformation parameter.'' 

\begin{figure}[h]
	\centering 
	\includegraphics[scale=0.42]{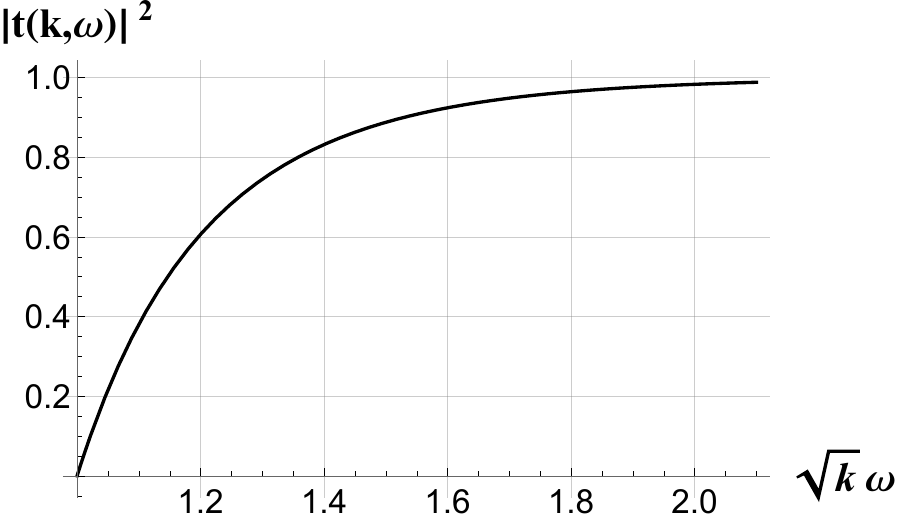}
	\caption{The transmission coefficient of a massless, spin-less probe sent from the left mouth of the wormhole to the right mouth as a function of $\sqrt{k} \omega$.}
		\label{Transmission} 
\end{figure}

      The action of the bosonic part of the deformed CFT can be re-expressed in terms of a non-linear sigma model
      \begin{align}
      	\label{tot} 
      	S_{\text{tot}} &=  		S_{\text{WZW},\text{SL}(2,\mathbb{R})_k}\Big[t-2H \sqrt{\frac{k_G}{k}}\varphi,\phi,\rho,\psi^1,\psi^2,\psi^3\Big]\nonumber\\
      	&~~+\frac{k_G(1+2H^2)}{4\pi} \int d^2z~ \partial \varphi \bar{\partial}\varphi+\frac{1}{2\pi}\int d^2 z ~ \psi_{\varphi} \bar{\partial} \psi_{\varphi}~~. 
      \end{align}
      Eq.~(\ref{tot}) is manifestly an action of a CFT$_2$. Constant $B$-field components in the directions $\varphi-t$ have been added in the last equation as well.
       The metric and $B$-field extracted from the action in Eq.~(\ref{tot}) are given by
	\begin{align}
		\label{GB}
		ds^2 &= \frac{k}{4}\Big[d\rho^2 + \cosh^2(\rho) d\phi^2 -(1+2H^2)\big(dt+\sinh(\rho)d\phi\big)^2\Big]+\frac{k_G}{4}\big(d\varphi + A\big)^2~~,\nonumber\\
		B^{(2)}&= -\frac{k}{4} \sinh(\rho) d\phi \wedge dt-\frac{1}{2}H \sqrt{k k_G} d\varphi \wedge dt~~,\nonumber\\
				A &= H \sqrt{\frac{2k}{k_G}} (dt+\sinh(\rho)d\phi)~~.
	\end{align}
	The coordinates have the following ranges: $-\infty<\rho < \infty$, $-\infty< t< \infty$, $0\leq \phi \leq 2\pi$ and $0\leq \varphi \leq 2\pi$.
	The ``three-dimensional'' metric that appears in the squared brackets on the R.H.S. of Eq.~(\ref{GB}) is referred to as ``squashed anti de-Sitter''.
Also, a $\text{U}(1)$ gauge field $A$ is generated from the Kaluza-Klein reduction along $\varphi$.\\
	In the third layer, one sends the deformation parameter to a specific imaginary value:
	\begin{equation}
		 H\to\frac{i}{\sqrt{2}}~~.
	\end{equation}
	Since the metric degenerates at this value of $H$, the limit can be taken carefully by  defining a rescaled time coordinate:
		\begin{equation}
		T = \sqrt{\frac{k}{4}\big(1+2H^2\big)}t~~.
	\end{equation}
	The range of $T$ is $-\infty<T<\infty$.
	The outcome of the third stage is a target space whose fields are given by
	\begin{align}
		\label{sol}
		ds^2 &= -dT^2 + \frac{k}{4} \Big(d\rho^2 +\cosh^2 (\rho)d\phi^2\Big)+\frac{k_G}{4}\big(d\varphi + A\big)^2~~,\nonumber\\
		B^{(2)}&=-\frac{k}{4} \sinh(\rho) d\phi \wedge dT~~,\nonumber\\
		A &= i \sqrt{\frac{k}{k_G}} \sinh(\rho)d\phi~~.
\end{align}
This imaginary field will be discussed towards the end of this subsection.
Upon a Kaluza-Klein reduction along $\varphi$, the target spacetime metric is that of a direct product of a Euclidean wormhole and Lorentzian time.  \\ An interesting quantity that probes the traversability of the wormhole in question is the transmission coefficient for string or brane excitations sent from the left mouth of the wormhole to the right mouth. In the regime $k\to \infty$ where a semiclassical approach is applicable, it can be calculated that a massless, spin-zero string probe experiences a  P\"oschl-Teller effective potential (up to an additive constant), which acts as a potential barrier. Solving for the transmission amplitude as a function of $\omega$, the frequency of the string, and taking zero angular momentum number $m=0$ for simplicity, one obtains
\begin{equation}
	\label{t}
	t(k,\omega) = \frac{\Gamma\big(\frac{1}{2}-i\mu\big)^2} {\Gamma(-i\mu)\Gamma(1-i\mu)}~~,~~ \mu\equiv  \frac{1}{2}\sqrt{  k\omega^2-1} ~~.
\end{equation}  
A plot of the transmission coefficient in the semiclassical regime using the absolute value square of Eq.~(\ref{t}), namely $\tanh^2(\pi \mu)$, appears in Fig.~\ref{Transmission}.\\
\newpage The conclusion from this calculation is that the wormhole becomes more traversable as the energy of the probe increases (though the energy should not exceed $\frac{1}{g_s^2}$ where $g_s$ is the string coupling, to stay in the probe approximation). Already at $\frac{2}{\sqrt{k}}\leq \omega$, the traversability of the perturbation through the wormhole is nearly perfect. 
It would be interesting to find a solution $t=t(k,\omega)$ that is valid for order-one $k$, in the spirit of the result for the reflection coefficient in the $\text{SL}(2,\mathbb{C})_k/\text{SU}(2)$ gauged WZW model~\cite{Teschner:1997ft}, in which the $k\to \infty$ reflection coefficient appears as a multiplicative factor. In a different context, the transmission coefficients of probes scattered in a traversable wormhole have been computed in~\cite{FFT}.  \\
Several comments are in order. First, despite the imaginary nature of the one-form $A$, the central charge of the worldsheet CFT in question is a positive real number. Note that a conventional example of a coset model based on the $\text{SL}(2,\mathbb{C})_k/\text{SU}(2)$ quotient admits an imaginary $B$-field, similarly to the imaginary $A$-field in the present context. Second, for large $k$, the square root of minus one in $A$ is crucial for the fields in Eq.~(\ref{sol}) to solve the supergravity equations, with the interpretation that the gauge field supports the wormhole. For the critical value of $H$, the stress-energy tensor appearing in Einstein's field equations is
\begin{equation}
	\label{T}
	T_{\mu \nu} = \frac{k_G}{4} F_{\mu \rho} F_{\nu} ^{~\rho} - \frac{k_G}{16} g_{\mu \nu} F_{\alpha \beta} F^{\alpha \beta}~. 
\end{equation}
Here, $g_{\mu \nu}$ is the string-frame target spacetime metric and $F=dA$. Natural units are used in writing Eq.~(\ref{T}).  Computing the energy density, one finds $T^{TT} = -\frac{1}{k}$, which violates the weak energy condition. Moreover, it can be verified that the null energy condition is violated as well. The matter source in question is a special condensate of heterotic closed string modes corresponding to an imaginary graviphoton mode $A$. The action of the effective field theory at large $k$ also has a negative cosmological constant, which is important for the solution to exist.
The gauge field and cosmological constant render the wormhole traversable for probes of sufficient energy~\footnote{A recent paper presented examples of traversable wormholes supported by imaginary-valued scalar fields is~\cite{Begines:2026mlv}.}. Third, the complex Lagrangian of the deformed theory with an imaginary $H$ implies that the worldsheet CFT in question is non-unitary (likewise, the $\text{SL}(2,\mathbb{C})_k/\text{SU}(2)$ CFT is non-unitary). The absence of unitary evolution in this CFT for the target space EAdS$_2$ wormhole is consistent with the loss of quantum coherence mediated by wormholes \cite{Hawking:1987mz}, \cite{Giddings:1987cg}. However, the particular cause of lost of quantum coherence in these references is that topology change in quantum gravity entails wormholes that connect a parent Universe to baby Universes\footnote{Quantum coherence loss occurs under the assumptions that (a) observers lack access to the baby Universes, (b) the quantum mechanics of baby Universes admits a non-trivial Hilbert space, and (c) the baby Universe is not in an $\alpha$-state where one may trade quantum coherence loss by uncertainty in coupling constants that appear in the action of the effective field theory~\cite{Coleman:1988cy}.}. This is not the reason for the loss of quantum coherence in the IKOP CFT with an imaginary deformation parameter. It should be pointed out that the CFTs in subsection~\ref{sec:sim} are reflection positive. Fourth, real, nonzero values of the deformation parameters $H$ bring about the pathology that closed timelike curves exist in the target spacetime. In contrast, for imaginary values of $H$, like $\frac{i}{\sqrt{2}}$, they are absent.  

In the special case $k=4$, where the AdS$_2$ length scale is comparable to the string scale, one can cancel the Weyl anomaly of the construction as follows. On the right-moving, supersymmetric sector, the right-movers' central charge is $c_R ^{(a)}=6 + \frac{6}{k} = \frac{15}{2}$. Taking the product with a free CFT of five chiral, compact bosons and five fermions adds another $c_R ^{(b)}=\frac{15}{2}$. The entire right-moving central charge is then $15$. On the left-moving sector, the initial central charge is $c_L ^{(a)} = 4+\frac{6}{k-2}=7$. Supplementing this by the $\text{SO}(32)$ or $E_8\times E_8$ chiral CFT provides $c_L ^{(b)}=16$, and by three more chiral, free, compact bosons with $c_L ^{(c)}=3$, the total left-moving central charge adds up to $c_L = 26$. Conventional $\{b,c\}$ and $\{\beta,\gamma\}$ ghost CFTs give rise to anomaly cancellation.  This suggests that the heterotic string theory admits an exact classical solution described by these products of CFTs. It would be interesting to verify this suggestion and whether or not the wormhole in question is stable. 
	\subsection{Double-Cone Wormholes from the Worldsheet} 
	\label{sec:double}
	Reference~\cite{Lowe:1994gt} demonstrated a three-stage procedure for obtaining a worldsheet CFT for string propagation in target spacetime containing two AdS$_2$ Rindler patches.~\footnote{An alternative method that results in this AdS$_2$ target spacetime with a gauge field is based on a quotient of the $\text{SL}(2,\mathbb{R})_k\times \text{U}(1)$ group manifold by another $\text{U}(1)$ subgroup~\cite{Giveon:2004zz} (a brief review appears in section 5 of~\cite{Betzios:2023jco}).} I apply it to generate a worldsheet description of a target space with a real double-cone wormhole in the Lorentzian signature.   
	
	In the first stage, the $\text{SL}(2,\mathbb{R})_k$ WZW model is considered, with group elements parametrized by the variables $t_L,t_R$ and $\rho$: 
		\begin{equation}
			\label{g}
		g(t_L,t_R,\rho) = \begin{pmatrix}
			\exp\big(\frac{t_L+t_R}{2}\big)\cosh\big(\frac{\rho}{2}\big)  & 			\exp\big(\frac{t_L-t_R}{2}\big)\sinh\big(\frac{\rho}{2}\big) \\
			\exp\big(\frac{t_R-t_L}{2}\big)\sinh\big(\frac{\rho}{2}\big)  & 			\exp\big(-\frac{t_L+t_R}{2}\big)\cosh\big(\frac{\rho}{2}\big) 
	\end{pmatrix}~~.
	\end{equation}
	The parameter $\rho$ ranges in $-\infty< \rho<\infty$, and $t_R$ and $t_L$ are also in the same range. 
	Plugging Eq.~(\ref{g}) into the WZW action in Eq.~(\ref{WZW}) yields 
	\begin{align}
		\label{SWZW2}
		S_{\text{WZW},\text{SL}(2,\mathbb{R})}[\rho,t_L,t_R]= \frac{k}{4\pi} \int_{\Sigma} d^2 z ~ \Big(\partial \rho \bar{\partial} \rho + \partial t_L \bar{\partial} t_L +\partial t_R \bar{\partial} t_R+2\cosh(r)\partial t_L\bar{\partial} t_R  \Big)~~.
	\end{align}
	The central charge of the CFT is
	\begin{equation}
		c = \frac{3k}{k-2}~~.
	\end{equation}
	  A constant dilaton, and the following three-dimensional metric and $B$-field describe the associated target spacetime
	\begin{align}
		ds_3 ^2&= \frac{k}{4} \Big(d\rho^2 + dt_L ^2 + dt_R ^2 +2\cosh(\rho) dt_L dt_R\Big)~~,\nonumber\\
			B^{(2)} &= \frac{k}{4}  \cosh(\rho) dt_L \wedge dt_R~~.
	\end{align}
	In the second stage, one quotients this theory by the integers subgroup of time translations $t_R$, which renders $t_R$ a compact dimension of periodicity $\frac{2\pi}{\gamma}$, where $\gamma$ is a real parameter (in other words, this is an orbifold operation). 
	
	In the third stage, one rewrites the three-dimensional metric in terms of the two-dimensional metric transverse to $t_R$:
	\begin{equation}
		ds^2 _3 = ds_2 ^2 + e^{2D} \big( dt_R + A_{\mu} dx^{\mu}\big)^2~~,
	\end{equation} 
	where $D$ is a ``radion'' field and $A$ is a gauge field. 	The range of $t_R$ is $[0,\frac{2\pi}{\gamma})$, with the other variables non-compact.
	 The outcome of the approach is an AdS$_2$ target spacetime with a Kaluza-Klein gauge field:
	\begin{align}
		ds_2^2 &= \frac{k}{4}\big(- \sinh^2 (\rho) dt_{L} ^2 +d\rho^2 \big)~~,~~ \nonumber\\
		A &= \gamma \cosh(\rho) dt_{L}~~,~~ \nonumber\\
				D &= \frac{1}{2} \log\Big(\frac{k}{4\gamma^2}\Big)~~.
	\end{align} 
	One can perform yet another stage: compactify $t_L$ on a circle of circumference $\widetilde{T}$ so that $0\leq t_L\leq \widetilde{T}$. The resulting target spacetime then contains a two-dimensional manifold with a double-cone wormhole where one cone is at $\rho>0$, the other cone is at $\rho<0$, and a contact point at $\rho=0$.
	 The torus worldsheet path integral is proportional to $\widetilde{T}$:
	 \begin{equation}
	 	\label{Z}
	 	Z_{\text{worldsheet}}(\widetilde{T},\tau,\bar{\tau}) = \int D t_L D t_R D \rho~ e^{-S_{\text{WZW},\text{SL}(2,\mathbb{R})/\mathbb{Z}}} \propto \widetilde{T}~~.
	 \end{equation}
	  The reason for the last transition is that one can separate the path integral into a zero mode of $t_L$ and nonzero modes of $t_L$, and integrating over the former produces the factor of $\widetilde{T}$. More precisely, the real-time partition function is defined by an analytical continuation of the conventional partition function, with the target space Euclidean, but this does not change the result. This genus-independent result gives rise to a ``ramp'' in the spectral form factor $Z(iT) Z(-iT)$ (for zero inverse temperature $\beta=0$ and boundary time $T$ proportional to $\widetilde{T}$) of a dual spacetime CFT$_2$ that lives on the conformal boundaries of the AdS spacetime, which is a variant of the exactly marginal deformation of the symmetric product orbifold CFT. However, the spectrum of physical string states in the worldsheet CFT Hilbert space is not governed by random matrix statistics; this is consistent with the proposition that black holes are chaotic systems since the worldsheet CFT does not capture black hole microstates. 
	  
	  Several additional remarks are written. First, when $k=4$, adding 20 more free compact bosons (and the $\{b,c\}$ ghost CFT) yields an exact bosonic string background.  Second, for $\beta>0$, the complexified, off-shell configuration should be described in the framework of string field theory, which is beyond the scope of this paper.  Third, the target spacetime is singular at the tip $\rho=0$. The singularity is resolved by going to the three-dimensional line element.  
	 Fourth, an alternative construction of a double cone from the worldsheet is based on compactifying the Lorentzian time field in the two-dimensional black hole with an asymptotically linear dilaton in~\cite{Witten:1991yr}. 
	 
	 Fifth, the double-cone wormhole construction is unstable due to strings wound around the compactified time $t_L$, which constitute tachyonic modes. To see this, one can use the formula for the mass-squared of winding modes in the context of Euclidean string theory on $\mathbb{R}^{25} \times S^1 _{\beta}$ for the bosonic string, for example: 
	 \begin{equation}
	 	m^2 = \frac{w^2 \beta^2}{4\pi^2 (\alpha')^2}-\frac{4}{\alpha'}~,
	 \end{equation}
	 where $w$ is the winding number and $\beta$ is the inverse temperature. Continuing $\beta \to i \tilde{T}$, one finds a tower of tachyonic modes. In the curved spacetime, the formula for the mass-squared is changed by replacing the $\beta^2$ factor by $|g_{t_Lt_L}|=\frac{k \alpha'}{4} \sinh^2 (\rho)$. Reference~\cite{Jorge} has argued for the stability of the double-cone configuration for brane nucleation using the DBI action; here, a different approach that employs a formula for the mass-squared of certain string modes is adopted to reach the opposite conclusion. The main lessons are that double-cone wormhole target spacetimes exist within the worldsheet formalism, but are unstable.          
	\subsection{Worldsheet Perspective on Einstein-Rosen Bridges}
	\label{sec:ER}
This subsection proposes two separate definitions of Einstein-Rosen bridges in the regime where supergravity is inapplicable and the string coupling is small. References~\cite{Jensen:2013ora},\cite{Chernicoff:2013iga} considered the opposite regime where supergravity is reliable, and showed that a pair of entangled quark and anti-quark in a strongly coupled supersymmetric Yang-Mills theory, has a bulk picture with a string connecting the pair such that the worldsheet geometry admits a wormhole interpretation  (in one Weyl frame). Reference~\cite{Jafferis:2021ywg} argued that the Lorentzian interpretation of the Fateev, Zamolodchikov, and Zamolodchikov (FZZ) duality  \cite{FZZ},\cite{Kazakov:2000pm} entails that an Einstein-Rosen bridge in the Hartle-Hawking state on the two-sided black hole side corresponds to a condensate of folded strings in the thermofield-double state on the Lorentzian FZZ dual side. Below, the goal is to provide new but more conservative descriptions of Einstein-Rosen bridges from a worldsheet perspective.

Examples of two- and three-dimensional black hole solutions which admit an exact CFT$_2$ description are the gauged WZW models based on the $\text{SL}(2,\mathbb{R})_k/\text{U}(1)$, $\text{SL}(2,\mathbb{R})_k/\mathbb{Z}_2 $, and $(\text{SL}(2,\mathbb{R})_k \times \text{U}(1)_L) / \text{U}(1)$ quotients. Constant time slices in the Lorentzian two-sided black hole target spacetime solutions constitute Einstein-Rosen bridges. At large $k$, they are described in the worldsheet by the classical solutions:
\begin{equation}
	t(z,\bar{z}) = \text{constant}~~.
\end{equation} 
At small $k$, the gauged WZW CFTs become strongly coupled; in particular, the classical description fails. 

Focusing on the $\text{SL}(2,\mathbb{R})_k/\text{U}(1)$ example, an  approach to describe ``stringy Einstein-Rosen bridges'' is to utilize a more suitable description for small $k$ - the Sine-Liouville CFT, or, its supersymmetric version,  $\mathcal{N}=2$ Liouville SCFT. The action description of Sine-Liouville is
\begin{equation}
S_{\text{SL}} = \frac{1}{4\pi} \int d^2 \sigma \sqrt{\gamma}\Bigg( \gamma^{ab}\partial_a \rho \partial_b \rho +\gamma^{ab}\partial_a \theta \partial_b \theta + 8\pi \lambda e^{-2b\rho } \cos\big(\sqrt{k} (\theta_L -\theta_R)\big)- Q R \rho \Bigg)~~.  
\end{equation}
Here, $\theta_L$ and $\theta_R$ are the left-moving and right-moving parts of the compact boson $\theta$, the parameters $b$ and $Q$ satisfy
\begin{equation}
	 b= \frac{1}{2} \sqrt{k-2}~~,~~ b(Q-b)+\frac{k}{4}=1~~.
\end{equation}
The coupling parameter $\lambda$ is not physical in and of itself; it can be rescaled by shifting the zero mode of the dilaton.
Euclidean time is related to $\theta$ through 
\begin{equation}
	\tau = \frac{\beta \sqrt{k}}{2\pi} \theta~~.
\end{equation}
Now, in the context of semiclassical gravity, the Hartle-Hawking procedure prepares an initial state in a system by gluing a Euclidean section of the cigar onto a Lorentzian $t=0$ slice. One can then evolve the state under Lorentzian time evolution. The Euclidean section can be sliced at the Euclidean time values $\tau=\pm \beta/2$. This can be done by selecting $\tau_L =\tau_R= \pm\frac{\beta}{4}$ where the cuts are made. Beyond the regime of semiclassical gravity, it is proposed that the dual of ``stringy Einstein-Rosen bridge'' is  the classical solutions $\tau=\pm  \beta/2$ of the Sine-Liouville CFT. Note that $\tau=\frac{\beta}{2}$ is not connected to $\tau=-\frac{\beta}{2}$ because the Euclidean time does not shrink. Time evolution in the Lorentzian target spacetime of these solutions creates the dual of future Einstein-Rosen bridges, though here the focus is on $t=0$.  Variants of the FZZ duality to other black holes (one of which has appeared in~\cite{Jafferis:2021ywg}) would allow for the same logic to define Einstein-Rosen bridges in the stringy regime in these other black holes.  

A different approach to define ``stringy Einstein-Rosen bridges'' avoids any mention of the FZZ duality. Consider a quantum state in which the field $t(z,\bar{z})$ is fixed at some time slice of the Lorentzian worldsheet denoted by ``$C$'': 
\begin{equation}
	|t(z,\bar{z})_{C}=\text{constant} \rangle~~.
\end{equation}
It is proposed that this quantum state defines an Einstein-Rosen bridge in the worldsheet CFTs for Lorentzian two-sided black holes. If the semiclassical approximation is valid for the black hole interior away from the singularity, the bridge in question is stable because gravity propagates perturbations living on a given bridge forward in time. If, by contrast, a two-sided black hole has firewalls~\cite{Almheiri:2012rt}, the usual notion of ER bridges fails. In particular, ~\cite{Itzhaki:2018glf} has argued that instant folded strings are created at the bridge, constituting an instability mechanism. While the extent to which the definitions written in this subsection are useful for solving problems in black hole physics is limited, the description of Einstein-Rosen bridges from a worldsheet perspective suggested here could be viewed as ``precise.''     
\subsection{Closed Universe/Wormhole Transition}
\label{sec:transition}
Reference~\cite{Israel:2004vv} showed that a conformal manifold exists, that contains the $\text{SU}(2)_k \times U(1)$ WZW model and worldsheet CFT for a wormhole target space, connecting them by an exactly marginal current-current deformation. A decoupled Lorentzian time $t(z,\bar{z})$ is added to the field content of these worldsheet CFTs, and then I propose a new interpretation of this slightly improved conformal manifold  - as mediating a transition between a closed cosmology and a  wormhole with a boundary which is a pair of disconnected manifolds. A brief review of this protocol is provided,   similar to the one in subsection~\ref{sec:AdS2}.     

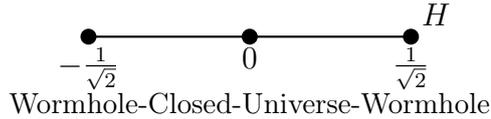
\begin{figure}[h]
	\begin{center}
	\begin{tikzpicture}[scale=3]
		
		\draw[thick] (-0.7071,0) -- (0.7071,0);

		\fill (-0.7071,0) circle (1pt);
		\fill (0,0) circle (1pt);
		\fill (0.7071,0) circle (1pt);
		
		\node[below] at (-0.7071,0) {$-\frac{1}{\sqrt{2}}$};
		\node[below] at (0,0) {$0$};
		\node[below] at (0.7071,0) {$\frac{1}{\sqrt{2}}$};
		
		\node[right] at (0.7071,0.1) {$H$};
		
		\node[below] at (0,-0.2) {\small Closed-Universe};
		\node[below] at (-0.75,-0.2) {\small Wormhole-};
		\node[below] at (0.75,-0.2) {\small -Wormhole};
	\end{tikzpicture}
\end{center}
\caption{A closed Universe phase exists in the bulk of the conformal manifold; its boundaries admit a wormhole interpretation. (Jokingly, the conformal manifold itself is a wormhole!)}
	\label{Fig1} 
\end{figure}

The action of the WZW model with the group manifold $\text{SU}(2)_k$ parametrized by $(\alpha,\beta,\gamma)$ (in Eq.~(\ref{para})) is
\begin{equation}
	S = \frac{k}{8\pi}\int d^2 z  \Big(\partial \alpha \bar{\partial} \alpha + \partial \beta \bar{\partial} \beta + \partial \gamma \bar{\partial} \gamma + 2\cos(\beta) \partial \alpha \bar{\partial} \gamma\Big)~~.
\end{equation}
One of the chiral currents of the model that will appear in an exactly marginal deformation of the CFT is
\begin{equation}
	J^3 = k\big(\partial \gamma + \cos(\beta) \partial \alpha\big)~~.
\end{equation}
A ``magnetic'' marginal deformation of the $\text{SU}(2)_k$ WZW CFT is
\begin{equation}
	 \delta S_{\text{magnetic}} = \frac{H}{2\pi} \sqrt{\frac{k_G}{k}} \int d^2 z ~ J^3 \bar{J}_G~~,
\end{equation}
where $\bar{J}_G$ is the current of a separate $U(1)$ WZW model. Following a Kaluza-Klein reduction about that $U(1)$ direction, the resulting target space is that of a squashed sphere:
\begin{align}
	ds^2 &= \frac{k}{4}\Big[d\beta^2 + \sin^2 (\beta) d\alpha^2 +(1-2H^2)\big(d\gamma+\cos(\beta) d\alpha\big)^2\ \Big]~,\nonumber\\
	A &= \sqrt{\frac{2k}{k_G}}H (d\gamma + \cos(\beta) d\alpha)~,\nonumber\\
	B^{(2)} &= \frac{k}{4} \cos(\beta) d\alpha \wedge d\gamma~~.
\end{align}
 The ranges of the Euler angles are $0\leq \alpha \leq 2\pi, 0\leq \beta \leq \pi$ and $0\leq \gamma \leq 2\pi$.
Now, suppose one takes the limit(s) $H\to \pm \frac{1}{\sqrt{2}}$. This yields a target space metric
\begin{equation}
	\label{S1S2} 
	ds^2 = dy^2+\frac{k}{4} \big(d\beta^2 + \sin^2 (\beta) d\alpha^2\big)~~,
\end{equation}
where a fixed, non-compact coordinate is $y$ defined in
\begin{equation}
	y = \sqrt{\frac{k}{2}\Big(\frac{1}{2}-H^2\Big)}\gamma~~.
\end{equation}
Decompactifying $\gamma$, the range of $y$ is $-\infty < y <\infty$.
Physically, the squashing of the sphere creates a hierarchy of scales between the fiber and the two-sphere base, with a limiting case of a wormhole when the fiber decompactifies.  
Now, one can add Lorentzian time to this family of CFTs parametrized by $H$. The point $H=0$ of the manifold admits an interpretation in terms of a closed cosmology. Also, the squashed sphere (times time) can also be interpreted as a ``closed Universe'' for any $H^2<\frac{1}{2}$. When $H\to \pm \frac{1}{\sqrt{2}}$ and $k$ is large, the CFT describes string propagation in a wormhole with two disconnected boundaries of the form $\mathbb{S}^2 \times \mathbb{R}_t$. See Fig.~\ref{Fig1}. 
\\
The main point of this subsection is that a transition exists between these target spaces when varying the parameter $H$. Note that a conventional time evolution in target spacetime or the worldsheet is different from tuning $H$, however, there exists a worldsheet CFT that generalizes the family above, whose target spacetime does interpolate between the two targets dynamically~\cite{Elitzur:2002vw}~\footnote{I thank Amit Giveon for pointing out this reference.}. 

A few general comments follow. First, one can interpret the endpoints of the conformal manifold as unstable to the magnetic deformation in the direction to the interior of the manifold. 
	
	Second, at the level $g_s\to 0$ (where $g_s$ is the string coupling) valid for the worldsheet approach utilized here, the dimension of the Hilbert space of the worldsheet CFT at $H=0$ with a closed cosmology target is infinite because states can be produced by acting with current operators $J^a _{-n}$ with any integer $n$ (and $a=1,2,3$) on highest-weight states. This result is also reproduced in quantum field theory on curved spacetime in the closed cosmology background. However, working at finite $g_s$ is expected to yield a finite Hilbert space dimension. This is predicted by a Euclidean string field theory path integral approach in a saddle-point approximation about the closed Euclidean target space, assuming it admits a Hilbert space interpretation. In fact, the on-shell, closed string field theory action vanishes $I_{\text{CSFT}}=0$ when the target space in question is compact \cite{Erler:2022agw}. This is further supported by \cite{Kazakov:2001pj},\cite{Brustein:2021cza},\cite{Chen:2021dsw} that demonstrated that the dilaton equation of motion implies that the on-shell action of a string theory background is a boundary term. Now, a modern phrasing of a proposal in~\cite{Gibbons:1976ue} is that the dimension of the Hilbert space of states in the static patch is given by the logarithm of the gravitational path integral over the Euclidean sphere. Adopting this proposal for the simpler closed cosmology $\mathbb{R}_t \times \text{SU}(2)_k$, it follows that the Hilbert space dimension receives no $\frac{1}{g_s^2}$ contribution. This conclusion is consistent with the claim that the dimension of the Hilbert space of baby Universes is one, which was argued from different perspectives in~\cite{Marolf:2020xie},\cite{McNamara:2020uza}.   
	
	Third, the transition described here between closed Universes and wormholes is qualitatively similar to the black hole/string transition, that was studied using the worldsheet formalism in~\cite{Giveon:2005mi}, where the entropy of the microcanonical ensemble of highly-excited states in $\text{SL}(2,\mathbb{R})_k/\text{U}(1)$ and $\text{SL}(2,\mathbb{R})_k/\mathbb{Z}_2$ theories was calculated as a function of the level $k$. 
	
	 I believe that the worldsheet approach offers additional lessons about closed Universes that cannot be inferred straightforwardly from a gravitational path integral approach. In particular, it would be interesting to revisit the Nappi-Witten worldsheet CFT for a closed, inhomogeneous Universe target spacetime~\cite{Nappi:1992kv}. 
	\section{Summary and Future Directions}
	\label{sec:final}
This paper introduced several constructions of worldsheet CFTs that describe string propagation in target-space wormholes. The worldsheet formalism allows one to discuss properties of wormhole targets in the stringy regime, using a language historically used to describe string propagation outside black holes. This goes beyond the state of the art in the field, which has focused on wormholes in the context of supergravity and the gravitational path integral. This work also provided a few exact classical string backgrounds with two disjoint boundaries connected together, following the rules of string theory. 

The simplest type of wormhole target spaces is a direct product of $\mathbb{R}$ and a compact manifold. Then, a worldsheet construction by~\cite{Israel:2004vv} was reviewed, which contains a Euclidean AdS$_2$ wormhole using a complex deformation of the $\text{SL}(2,\mathbb{R})_k \times U(1)_{k_G}$ WZW model. The next example reviewed in this paper is based on an orbifold of $\text{SL}(2,\mathbb{R})_k$ CFT~\cite{Lowe:1994gt}, which gives rise to a pair of Rindler patches of AdS$_2$ spacetime, which, upon compactifying a timelike field, admits a double-cone interpretation. Einstein-Rosen bridges in the stringy regime were proposed for the $\text{SL}(2,\mathbb{R})_k/\text{U}(1)$ CFT in terms of FZZ dual classical solutions.  Finally, a conformal manifold parametrized by the coupling of an exactly marginal deformation of the $\text{SU}(2)_k$ WZW model was interpreted as allowing one to transform a closed Universe target spacetime into a wormhole by an extreme squashing.

One basic future direction is to construct more examples of worldsheet CFTs with non-trivial, warped wormhole targets. It would be interesting to avoid the need for Kaluza-Klein reductions, or any complex marginal deformations, to achieve simple wormhole target spaces. If the logic of General Relativity coupled to matter fields extends to the worldsheet approach, then to fulfill this objective, one requires negative-energy sources that backreact on the geometry; an instantiation of such a source is instant folded strings \cite{Itzhaki:2018glf},\cite{Attali:2018goq}-\cite{Itzhaki:2025gdv}. Negative-tension orientifold planes with RR fluxes and Casimir energy constitute other examples; however, it is not direct to describe them in the framework of the worldsheet formalism.\\ Another possible route for finding more examples of Euclidean wormhole target spaces is to patch pairs of T-dual spaces along a self-dual radius of the manifolds. For instance, in two dimensions, imagine gluing the cigar target space to the trumpet target space at a ``stretched horizon'' where the thermal circle is of order the string scale. An issue with this particular setup is an order-one string coupling near the origin of the trumpet; this problem also arises when attempting to interpret the Sine-Liouville CFT as describing a wormhole target. \\ Additionally, the quotient of the boundary manifold of EAdS$_3$ by a Fuchsian group $\Gamma$ used in~\cite{MaldacenaMaoz} for demonstrating that supergravity admits Euclidean wormhole solutions with EAdS$_3$ asymptotics, is potentially relevant to generating a worldsheet CFT for the pure-NS Maldacena-Maoz wormhole, an $\big(\text{SL}(2,\mathbb{C})_k/\text{SU}(2)\big)/ \Gamma$ gauged WZW model~\cite{Troost}. 

An orthogonal future direction is to study how incorporating the Schwarzian degree of freedom into wormhole constructions with AdS$_2$ geometries, which represents quantum fluctuations in the throat length, affects the stability and traversability properties of these targets.

In addition, it is intriguing to study the relationship between stringy wormholes and non-local interactions, either in the exact theory or in an effective field theory thereof, and to determine the distance scales associated with such non-local interactions. While interactions in these CFTs are local on the worldsheet, integrating out long string degrees of freedom in the spacetime theory generates non-local interactions in target space.

Finally, it would be interesting to revisit the ER=EPR conjecture~\cite{Maldacena:2013xja} using the constructions of this paper, testing whether stringy wormholes correspond to quantum entanglement.  The mutual information associated with two of the wormhole's disconnected boundaries could serve as a useful quantity to test the relationship between the quantum entanglement of the degrees of freedom residing there and the wormhole's connectivity.
\subsection*{Acknowledgements}
I am grateful to Ofer Aharony, Panos Betzios, Roberto Emparan, Barak Gabai, Amit Giveon, Matthew Heydeman, Sunny Itzhaki, David Kutasov, Yaron Oz, Amit Sever, Marija Toma\v{s}evi\'{c}, and especially Jan Troost for discussions. Also, I thank Dan Isra\"el for a correspondence.  I thank Paris-Saclay Universit\'e  and Universitat de Barcelona for their hospitality. 
YZ is supported by the Israel Science Foundation, grant number 1099/24.

\end{document}